\documentclass[12pt, prd, aps, superscriptaddress, notitlepage, preprint]{revtex4-2}

\usepackage{natbib}
\usepackage{mathrsfs}
\usepackage[utf8]{inputenc}
\usepackage{amsfonts,bm}
\usepackage{amsmath,amssymb,graphicx,textcomp,xcolor}
\usepackage{hyperref}
\usepackage[caption=false]{subfig}
\usepackage{slashed}

\newcommand{\Aext}{\mathcal{A}^{\rm ext}}

\newcommand{\lla}{\left\langle}
\newcommand{\rra}{\right\rangle}
\newcommand{\tr}{\mathrm{Tr}}
\newcommand{\sinc}{\sin_{\rm c}}

\begin{document}

\newcommand*{\MIT}{Center for Theoretical Physics -- A Leinweber Institute, Massachusetts Institute of Technology, Cambridge, MA 02139, USA}\affiliation{\MIT}

\title{Vacuum polarization in QED with an electromagnetic background from the lattice}
\author{En-Hung Chao}\affiliation{\MIT}
\preprint{MIT-CTP/6089}
\date{August 23, 2026}

\begin{abstract}
We propose a framework to determine the vacuum polarization functions in QED in the presence of an electromagnetic background field in the spacelike region on the lattice.
This method consists in reweighting lattice Monte Carlo data generated without the background with the fermion determinant ratio evaluated in the continuous spacetime using worldline formalism.
This proposal can be further extended to other applications in lattice gauge theory, such as QCD in finite chemical potential.
\end{abstract}

\maketitle
\tableofcontents
\newpage

\section{Introduction}
Developments in experimental infrastructure with intense laser have driven studies of QED in the presence of external electromagnetic field (see Refs.~\cite{Fedotov:2022ely, Hattori:2023egw} for review).
As the intensity of the external fields increases, the fine-structure constant $\alpha\equiv e^2/4\pi$ is no longer a good expansion parameter for perturbative treatments. 
Instead, conjectured by Ritus and Narozhny~\cite{Ritus:1972ky, Narozhnyi:1980dc}, the parameter $g\equiv \alpha\chi^{2/3}$ with the dynamical quantum parameter $\chi=(e/m^3)\sqrt{-(F_{\mu\nu}p^\nu)^2}$, where $F_{\mu\nu}$ is the field strength, $p$ is a particle momentum and $m$ is the mass of the electron, might become the appropriate parameter.
Understanding this emergent feature of QED is important as future experiments with high-intensity laser will operate at $\chi\gtrsim 1$~\cite{Ur:2015cha, LUXE:2023crk, Borysov:2025ehq}.
Various theoretical works have attempted to verify this conjecture~\cite{DiPiazza:2020kze, Mironov:2021ohk, Heinzl:2021mji}.
In particular, in Ref.~\cite{Mironov:2020gbi}, the authors have performed calculations with certain types of radiative correction diagrams resummed to all order and provided evidence for the relevance of the parameter $g$ in a constant crossed field (CCF) configuration.
In spite of the significant endeavor and progress from analytic QED, it would be interesting to investigate the conjecture using non-perturbative methods where all radiative correction diagrams can be systematically included.  

Lattice gauge theory, where dynamical gauge fields are stochastically sampled from Markov Chain Monte Carlo simulations according to a Euclidean action, would be a suitable framework for this purpose.
Nonetheless, despite the success in simulating QED in the presence of external magnetic field on the lattice~\cite{Kogut:2023ruw, Sinclair:2025ber}, including a real, physical electric field in a dynamical simulation is not as straightforward due to the non-trivial phase of the Euclidean action when analytically continuing from Minkowski space.

Computing observables under complex actions is at the frontier of modern research activities in lattice gauge theory.
One idea of such is \textit{reweighting}: starting with a theory which can be simulated using Monte Carlo techniques, one amends the observable by including the correction from the sampled theory to the target theory in it~\cite{Fodor:2001pe}. 
A well-known example of application is QCD at high temperature and low chemical potential, which has been studied in Ref.~\cite{Borsanyi:2021sxv} with the reweighting factor obtained from a Taylor expansion in the ratio of the chemical potential to the temperature.
A significant advantage of such a method is that there is no need to generate new gauge ensembles as one performs a scan over external potentials.

Nevertheless, due to the imperfect overlap between both theories, the signal of the reweighted data degrades exponentially as the system size increases.
It is thus expected that reweighting-based techniques will fail to reproduce statistically meaningful results at some point.
Beyond conventional importance-sampling based Monte Carlo techniques, recent promising proposals for simulating systems with complex actions include Complex Langevin Equation~\cite{Berges:2005yt, Aarts:2008rr} and Lefschetz thimble~\cite{Cristoforetti:2012su}.
See Refs.~\cite{Nagata:2021ugx, Aarts:2026uiu} for more exhaustive lists of references for recent works.

Assuming that the overlap between the sampled and target theories is sufficiently good -- which should be the case as the free energy under a CCF configuration in the zeroth order in $\alpha$ is the same as that of the free theory~\cite{Schwinger:1951nm} -- the goal of the present work is to provide an alternative reweighting-based framework to study systems with a complex action due to fermion bilinears from first principles, which bypasses the truncated summation in the Taylor expansion approach and allows to probe a wider range of external potentials.

We now proceed with a brief preview of our proposal.
Our approach is inspired by the plethora of available analytic methods for computing fermion determinants -- or the one-fermion-loop effective actions -- at a fixed background field in the continuous space~\cite{Dunne:2007rt}.
As input, we utilize gauge configurations generated from lattice simulations without external potential.
We first apply a smoother to the lattice data with the high-frequency modes above a cutoff $\Lambda$ filtered out, so that one can rigorously interpolate them between the existing points on the spacetime grid. 
A convenient choice for the gauge smoother is gradient flow~\cite{Narayanan:2006rf, Luscher:2010iy}, which, at finite flow time $t=1/\Lambda^2$, suppresses the high-frequency modes exponentially.
After that is done, we can apply our method of choice to calculate the fermion determinant ratio semi-analytically from the interpolated lattice gauge fields with and without external potential to get the reweighting factor. 
Since the gradient-flowed data are band-width limited, the Whittaker-Kotel'nikov-Shannon interpolation theorem, which perfectly reconstructs the spectrum from the data on a grid, can be applied.
As the flow time has a mass dimension of minus two, we propose to calculate the fermion determinants using worldline formalism~\cite{Bern:1991aq, Strassler:1992zr}, under which the fermion determinants can be consistently regulated with a heat-kernel regularization at a proper time $t_h = t$.
To get the final physical result, we need to take the lattice spacing $a\to 0$ and $t\to 0$ limits in turn, which might require additional counter terms for renormalization.
An appealing feature of gradient flow for this application is that a flowed lattice gauge field does not require wave function renormalization~\cite{Luscher:2011bx}, rendering the overall renormalization in our proposal much simplified.

This paper is organized as follows.
In Section~\ref{sec:formalism}, the theory foundations of our proposed framework are explained in detail, with special focus on the application to QED in the presence of background fields.
In Section~\ref{sec:vp-ccf}, we show how to obtain the vacuum polarization functions in QED under a constant-crossed-field background in the spacelike region using our method.
In particular, we discuss the subtleties while connecting our Euclidean-space setup to the real-world physics in Minkowski space. 
Concluding remarks are made in Section~\ref{sec:concl}.
A brief review on the Seeley-DeWitt coefficients, relevant for the renormalization in our framework, is given in Appendix~\ref{sec:seeley-dewitt}.
Finally, a numerical study of the large-separation behavior of the current-current correlator in free-field QED with external electric and magnetic fields is presented in Appendix~\ref{sec:free-field}. 

Unless otherwise stated, we will be referring to quantities in Euclidean space.

\section{Reweighting with worldline formalism}\label{sec:formalism}

Consider a system described by the partition function
\begin{equation}\label{eq:zpar}
\mathcal{Z}[\Aext] = 
\int\mathcal{D}[A,\psi,\bar{\psi}]
e^{-S_{\rm g}[A] - S_{\rm f}[\Aext, A,\psi,\bar{\psi}]}\,,
\end{equation}
where $S_{\rm g}$ is the kinetic action for the dynamical gauge field $A$, $\Aext$ is an external, static vector potential, which can be complex-valued, and 
\begin{equation}\label{eq:sferm}
S_{\rm f}[\Aext, A,\psi,\bar{\psi}] \equiv \bar{\psi} \slashed{D}[A]\psi + \bar{\psi}\slashed{\Aext} \psi\,
\end{equation}
with fermion fields $\psi$ and $\bar{\psi}$.
We work with Hermitian $\gamma$-matrices and the Dirac operator 
\begin{equation}
\slashed{D} \equiv \slashed{\partial} + i\slashed{A}+m
\end{equation}
satisfies $\gamma_5$-Hermiticity. 
Note that our convention is such that the coupling constant is absorbed in the gauge field as commonly done in lattice gauge theory literature.

For simplicity, we limit ourselves to one fermion species. 
Integrating out the fermionic degrees of freedom in Eq.~\eqref{eq:zpar} gives
\begin{equation}\label{eq:zpar-noferm}
\begin{split}
\mathcal{Z}[\Aext] = & \int \mathcal{D}[A]
e^{-S_{\rm g}[A]-\Gamma[A,0]}\;\mathcal{V}[A,\Aext]
= 
\lla \mathcal{V}[A,\Aext] \rra
\,,
\end{split}
\end{equation}
where the reweighting factor is defined via
\begin{equation}
\mathcal{V}[A,\Aext] \equiv
\frac{\det\left( \slashed{D}[A] + \slashed{\Aext} \right)}{\det \slashed{D}[A] }
=
 e^{-\Delta \Gamma[A,\Aext]}\,,
\end{equation}
with
\begin{equation}\label{eq:lndet}
\Delta\Gamma[A,\Aext] =
\Gamma[A,\Aext] - \Gamma[A,0]\,,
\quad
\Gamma[A,\Aext] \equiv 
-\ln\det\left( \slashed{D}[A] + \slashed{\Aext} \right)\,.
\end{equation}
Note that, in the presence of an external potential, a pure-gauge observable will simply need to be reweighted by the factor $\mathcal{V}$; whereas for a fermionic observable, we need to include the external potential in the Dirac equation while computing the propagator additionally.

In the following, we will first briefly review the theoretical foundations of the two techniques involved in our proposal, namely \textit{gradient flow} and \textit{worldline formalism}, in Section~\ref{sec:wf} and \ref{sec:wl} respectively.
Then, our framework and the workflow including the renormalization procedure to study QED in an external electromagnetic (E\&M) field will be discussed in Section~\ref{sec:method}.

\subsection{Gradient flow}\label{sec:wf}
We will adopt the notations from Ref.~\cite{Luscher:2010iy} specifically for the discussion in this subsection.

Introduce a flow-time variable $t>0$. 
We allow the lattice gauge field $A_\mu$ to evolve in the fictitious flow-time direction in a gauge-covariant way according to
\begin{equation}
\frac{\partial}{\partial t} B_\mu = D_\nu G_{\nu\mu}\,,\quad B_\mu\mid_{t=0} = A_\mu\,,\quad
G_{\mu\nu} = \partial_\mu B_\nu - \partial_\nu B_\mu + 
[B_\mu,B_\nu]\,.
\end{equation}
where $B_\mu(t)$ is referred to as the flowed gauge field and
\begin{equation}
D_\mu = \partial_\mu+ [B_\mu,\cdot]\,,
\end{equation}
is the associated covariant derivative.

At finite flow times $t$, the UV fluctuations of a Fourier mode of a flowed gauge field associated with momentum $p$ are $e^{-p^2t}$ suppressed.
As a consequence, no wave function renormalization is needed for the flowed gauge field~\cite{Luscher:2011bx}.
In lattice QCD, gradient flow is often used for scale setting~\cite{Luscher:2010iy, BMW:2012hcm} and can be similarly generalized to the case of fermions and composite operators, with examples of application in parton distribution functions~\cite{Monahan:2016bvm, Shindler:2023xpd} and lifetimes of mesons~\cite{Black:2023vju, Black:2024iwb}.
A summary of recent developments in this field can be found in Ref.~\cite{Shindler:2022tlx}.

In the remainder of the paper, gradient flow is only applied to the gauge field when referred to. 

\subsection{Worldline formalism}\label{sec:wl}

The worldline formalism first appeared in the context of first quantization proposed by Feynman~\cite{Feynman:1950ir} and was later revived in scattering amplitude~\cite{Bern:1991aq} and effective action calculations~\cite{Strassler:1992zr}.
The idea is to first write the fermion determinant in a heat-kernel/proper-time regularization.
At a fixed proper-time $T$, quantum fluctuations are encoded in a path integral of spacetime coordinates, plus spin and color degrees of freedom, if present.
For a pedagogical derivation of the formalism and its applications, we refer the reader to Ref.~\cite{Schubert:2001he}.

For illustration purposes, we will focus on the case of spinor QED, whose effective action can be written as
\begin{equation}\label{eq:ea-spin}
	\Gamma[A] \equiv -\ln\det\slashed{D}[A] =  -\frac{1}{2}\int^\infty_0 \frac{dT}{T} e^{-m^2T}\int\mathcal{D}x(\tau)\mathcal{D}\psi(\tau) e^{-\int^T_0 d\tau L} \,,
\end{equation}
where the coordinates $x_\mu$ treated as dynamical fields and the fermionic field $\psi$ satisfy periodic and anti-periodic boundary conditions in the worldline coordinate $\tau$ respectively, and the worldline Lagrangian is defined as
\begin{equation}
L = \frac{1}{4}\dot{x}_\mu\dot{x}^\mu + \frac{1}{2}\psi_\mu \dot{\psi}^\mu + i\dot{x}^\mu A_\mu - i\psi^\mu F_{\mu\nu} \psi^\nu\,,\quad
\dot{x}\equiv \frac{dx(\tau)}{d\tau}\,.
\end{equation}
Here $F_{\mu\nu}$ is the field strength of the gauge field $A_\mu$.

For numerical applications, it is more convenient to express the ``spin factor'' as a path-ordered integral~\cite{Schneider:2018huk}
\begin{equation}\label{eq:phispin}
\begin{split}
\Phi[x,F] \equiv& \int\mathcal{D}\psi\exp\left(- \int_0^T d\tau \left[
\frac{1}{2}\psi_\mu \dot{\psi}^\mu - i\psi^\mu F_{\mu\nu} \psi^\nu\right]\right)
\\ 
= &\quad \tr\mathcal{P}\exp
\left[
\frac{i}{2}\sigma^{\mu\nu}\int^T_0 d\tau
F_{\mu\nu} (x(\tau))
\right]\,,
\end{split}
\end{equation}
where the trace is taken over the spinor indices.
With this, we can rewrite Eq.~\eqref{eq:ea-spin} as
\begin{equation}\label{eq:gamma-qed}
\Gamma[A] = -\frac{1}{2}\int^\infty_0 \frac{dT}{T} e^{-m^2T}\int\mathcal{D}x\; \Phi[x,F]\mathcal{W}[x,A] e^{-\frac{1}{4}\int^T_0 d\tau\; \dot{x}^2} \,,
\end{equation}
where 
\begin{equation}
\mathcal{W}[x,A]\equiv e^{-i\int_0^T d\tau \dot{x}^\mu A_\mu(x(\tau))}\,,
\end{equation}
is a Wilson loop.

Note that the integrand in the effective action in Eq.~\eqref{eq:gamma-qed} is divergent at the proper time $T\to 0$, so one needs to include counter terms in order to obtain a renormalized quantity. 
In practice, we can regulate the theory by imposing a cutoff $t_h$ as the lower bound of the $T$-integral, ie. 
\begin{equation}\label{eq:uv-reg}
\int^\infty_0 \frac{dT}{T} = \lim_{t_h\to 0}\int^\infty_{t_h}\frac{dT}{T}\,.
\end{equation}

This discussion can be generalized to other gauge theories with fermions.
In the case of QCD, one needs to take an extra color trace of the product $\Phi[x,F]\mathcal{W}[x,A]$.

\subsection{Proposed framework}\label{sec:method}
Our goal is to numerically evaluate the vacuum expectation value of an observable $\mathcal{O}$ under the partition function Eq.~\eqref{eq:zpar-noferm} via the ratio
\begin{equation}\label{eq:master}
\frac{1}{\mathcal{Z}[\Aext]}\lla \mathcal{O} \rra_{\mathcal{Z}[\Aext]} = 
\frac{\lla \mathcal{O} \mathcal{V}\rra}{\lla \mathcal{V}\rra}\equiv
\mathcal{R}_{\mathcal{O}}
\end{equation}
where the expectation values on in the middle are evaluated under the unperturbed action with partition function $\mathcal{Z}[0]$.
Our computation strategy is as follows.
Given a lattice gauge configuration $A^{\rm lat}$, we evolve it with gradient flow up to a flow time $t$. 
As the noise from the high-frequency part is filtered out, it enables a smooth interpolation between data point.

A good choice of interpolation scheme is the generalized version of the Whittaker-Kotel'nikov-Shannon sampling theorem in $N$ dimensions~\cite{
Whittaker1915, Kotelnikov2001, Whittaker1935, PETERSEN1962279}: for any function $f$ which in frequency domain has support contained in $[-1/2a,1/2a]^N$, the interpolation
\begin{equation}\label{eq:whittaker}
\tilde{f}(\mathbf{x}) = \sum_{\mathbf{n}\in\mathbb{Z}^N} f(a\mathbf{n})\prod^N_{i=1}\sinc\left(\frac{x_i-an_i}{a}\right)\,,\quad\forall\mathbf{x}\in \mathbb{R}^N \,,\,,
\end{equation}
with $\sinc t\equiv \sin \pi t/(\pi t)$, preserves the spectrum of the function $f$.
As the gradient flow further suppresses the high-frequency fluctuations in addition to the lattice regularization, we expect Eq.~\eqref{eq:whittaker} to work well and that only a few data points neighboring $\mathbf{x}$ would be needed for the series to converge.

Note that Eq.~\eqref{eq:whittaker} requires the function to live in a flat space. 
In the case of a gauge theory, one would need to first gauge transform the field $A^{\rm lat}$ to a fixed gauge orbit before applying Eq.~\eqref{eq:whittaker}, and then transform back after the interpolation. 
As Eq.~\eqref{eq:whittaker} preserves the spectrum, choosing a gauge orbit with gauge fixing condition local to the Fourier transform of $A^{\rm lat}$ will be a consistent scheme.

We then apply the worldline formalism to calculate the difference in effective action Eq.~\eqref{eq:lndet} with the interpolated field $\tilde{A}(t|a)$ from $A^{\rm lat}$ as input at a finite lattice spacing $a$.
Since gradient-flowed gauge fields do not require wave function renormalization in our convention, $\tilde{A}(t|a)$ can be smoothly extrapolated to its counter part in the continuum, $\tilde{A}(t)$.

As mentioned earlier, we impose a lower limit $t_h$ to regulate the UV divergence in the fermion determinant evaluation. 
Because of their mass dimensions, we can in fact set $t_h=t$ and have both regulator lifted by taking the $t\to 0$ limit to get the physical result.  
Here, the flowed gauge field is only used for the reweighting factor $\mathcal{V}$; the vacuum expectation values of Eq.~\eqref{eq:master} are evaluated under the unperturbed action with un-flowed gauge fields.
We arrive at the $t$-regulated reweighting factor
\begin{equation}\label{eq:t-reg-V}
\mathcal{V}(t,a) = \exp\left(-\Delta\Gamma_t[\tilde{A}(t|a),\Aext]\right)\,,
\end{equation}
where the subscript $t$ indicates that we have replaced the proper-time integral in the effective actions by their $t$-regulated version.

Keeping track of all the regulators, we rewrite Eq.~\eqref{eq:master} as 
\begin{equation}\label{eq:R-reg}
\mathcal{R}_{\mathcal{O}} = 
\lim_{t\to 0}\lim_{a\to 0}\left[ \frac{\lla \mathcal{O}^{\rm lat}[A^{\rm lat}](a)\mathcal{V}(t,a)\rra}{\lla \mathcal{V}(t,a) \rra}\right]^{\rm R}\,,
\end{equation}
where the superscript `R' implies that the quantity is renormalized.
Note that we take the $a\to 0$ prior to the $t\to 0$ limit.

As the operator $\mathcal{O}$ does not involve gradient flow, its renormalization follows the standard treatments in lattice gauge theory.
In the following, we will discuss the removal of the $t$-dependent divergences and the renormalization, with spinor QED in an E\&M background ´as an example.

\subsubsection{The small proper-time expansion}\label{sec:sptx}

Consider a heat-kernel regulated functional determinant 
\begin{equation}
\ln \det_t \mathcal{H}[A] \equiv -\int_t^\infty \frac{dT}{T} \text{Tr}\left( e^{-T\mathcal{H}[A]} \right)\,,
\end{equation}
where the operator $\mathcal{H}$ is defined via
\begin{equation}\label{eq:hheat}
\mathcal{H}[A] = -\mathcal{D}[A]^\mu \mathcal{D}[A]_\mu - \mathcal{E}[A]\,,\quad
\mathcal{D}_\mu \equiv \partial_\mu + iA_\mu\,,
\end{equation}
with $\mathcal{E}$ a potential term.

Upon quantization, the traced quantity becomes precisely the path integral in the proper-time parameterization in Eq.~\eqref{eq:ea-spin}.
This determinant can be written in terms of the Seeley-DeWitt coefficients of the operator $\mathcal{H}$ in four dimensions~\cite{DeWitt:1964mxt, Seeley:1967ea} (cf. Appendix~\ref{sec:seeley-dewitt} for details)
\begin{equation}
\ln \det_t \mathcal{H}[A] = -\frac{1}{16\pi^2} \int d^4x \left[ \frac{\text{Tr}(a_0)}{2t^2} + \frac{\text{Tr}(a_1)}{t} + \text{Tr}(a_2)\ln\left(\frac{1}{t}\right) \right] + \text{O}(t^0)\,.
\end{equation}

As a result, the small-$t$ behavior of the difference in effective actions in Eq.~\eqref{eq:t-reg-V} in the continuum limit can formally be written as
\begin{equation}\label{eq:sftx-gamma}
\begin{split}
\Delta_t\Gamma[\tilde{A}(t|a),\Aext] = & -\frac{1}{2}\Big(
\ln\det \mathcal{H}[A(t)-i\Aext] - \ln\det \mathcal{H}[A(t)]\Big)
\\
\xrightarrow{t \to 0} & \frac{1}{32\pi^2} \left[ \frac{1}{t} \int d^4x \, \text{Tr}(\Delta a_1) + \ln\left(\frac{1}{t}\right) \int d^4x \, \text{Tr}(\Delta a_2) \right] + \mathrm{O}(t^0)\,,
\end{split}
\end{equation}
where $\Delta a_1$ and $\Delta a_2$ are the differences between the Seeley-DeWitt coefficients $a_{1,2}$ calculated with and without the external gauge potential.

\subsubsection{Renormalization of QED in an E\&M background}
In this context, we have, for the potential term in Eq.~\eqref{eq:hheat},
\begin{align}
\mathcal{E}[A] &= \frac{i}{2}\sigma^{\mu\nu}F_{\mu\nu} - m^2\,, \\
\mathcal{E}[A-i\Aext] &= 
\frac{i}{2}\sigma^{\mu\nu}\left(F_{\mu\nu} + \mathcal{F}^{\rm ext}_{\mu\nu}\right) - m^2\,,
\end{align}
where we introduce the field strength of the external potential via
\begin{equation}
\mathcal{F}^{\rm ext}_{\mu\nu}\equiv -i (\partial_\mu \Aext_\nu - \partial_\nu \Aext_\mu)\,.
\end{equation}
Note the unusual factor of $i$ due to our convention for $\Aext$ [Eq.~\eqref{eq:sferm}].

Using the results from Appendix~\ref{sec:seeley-dewitt}, we obtain
\begin{equation}
\tr(\Delta a_1) = 0 \,,\quad
\tr(\Delta a_2) 
= -\frac{2}{3}\left(2\tilde{F}^{\mu\nu}(t|a)\mathcal{F}^{\text{ext}}_{\mu\nu} + (\mathcal{F}^{\rm ext}_{\mu\nu})^2\right) \,,
\end{equation}
where $\tilde{F}^{\mu\nu}(t|a)$ is the field strength of $\tilde{A}^{\mu}(t|a)$.
As a result of the smoothness of the flowed gauge field, we can take the $a\to 0$ limit at fixed $t>0$ and obtain
\begin{equation}\label{eq:sftx-sfqed}
\lim_{a\to 0}\Delta\Gamma_t[\tilde{A}(t|a),\Aext] \underset{t \to 0}{\sim}-\frac{1}{32\pi^2} \ln\left(\frac{1}{t}\right) \int d^4x \left[ \frac{2}{3}(\mathcal{F}_{\mu\nu}^{\rm ext})^2 + \frac{4}{3}\tilde{F}^{\mu\nu}(t)\mathcal{F}^{\rm ext}_{\mu\nu} \right] + \mathrm{O}(t^0)\,,
\end{equation}
where $\tilde{F}^{\mu\nu}(t)$ is the field strength of $\tilde{A}^{\mu}(t)$.

The mixing between the external and the dynamical fields makes appear logarithmic divergence in the $t\to 0$ limit which cannot be factorized out and needs to be handled on a configuration-by-configuration basis before taking the $a\to 0$ limit. 
Such a divergent mixing is generated at the lowest level by the fermion loop coupling at the same time to both the external field and the gauge field.
In the convention adopted for Eq.~\eqref{eq:sftx-sfqed}, we absorb the bare charge in the definition of the vector potential. 
While the gradient-flowed field strength $F_{\mu\nu}(t)$ does not require any further renormalization, the bare external field $\mathcal{F}_{\mu\nu}^{\rm ext}$ needs to be multiplicatively renormalized.

We define the wave-function renormalization constant $Z_{\rm ext}$ to connect the bare with the gradient-flow renormalized quantity at a scale $\mu$
\begin{equation}
\mathcal{F}_{\mu\nu}^{\rm ext} = Z_{\rm ext}(\mu,t) \mathcal{F}_{\mu\nu}^{\rm ext, R}(\mu) = (1 + \delta Z_{\rm ext}) \mathcal{F}_{\mu\nu}^{\rm ext, R}(\mu)\,.
\end{equation}
Substituting this back into the effective action, and demanding that $\delta Z_{\rm ext}$ exactly cancels the logarithmically divergent piece to render the effective action finite, we set:
\begin{equation}
\delta Z_{\rm ext}(\mu,t) = -\frac{1}{24\pi^2} \ln(\mu^2 t)\,.
\end{equation}
Thus, the renormalization constant reads:
\begin{equation}
Z_{\rm ext}(\mu,t) = 1 - \frac{1}{24\pi^2} \ln(\mu^2 t)\,.
\end{equation}
We define this as the \textit{gradient-flow minimal subtraction} (GFMS) scheme.
Note that $Z_{\rm ext}(\mu, t)$ contains exactly the divergence in one-loop beta function in QED, which comes with no surprise as gauge invariance imposes the renormalization condition for the fine-structure constant $\alpha = \alpha_0 Z_{\rm ext}$ from its bare value.

To summarize, we renormalize Eq.~\eqref{eq:sftx-sfqed} by adding a counter term, yielding
\begin{equation}
\Delta\Gamma_t^{\rm 1-loop, R}[\tilde{A}(t|a),\Aext]\equiv
\Delta\Gamma_t[\tilde{A}(t|a),\Aext{}^{\rm ,R}(\mu)] + \delta Z_{\rm ext}(\mu,t) \int d^4 x \tilde{F}_{\mu\nu}(t|a) \mathcal{F}^{\rm ext, R}_{\mu\nu}(\mu)\,,
\end{equation}
calculated configuration-by-configuration.

\section{QED vacuum polarization tensor in constant crossed fields}\label{sec:vp-ccf}
In this section, we discuss how the vacuum polarization tensor can be calculated to all-order in presence of constant crossed fields, ie. when external constant electric and magnetic fields satisfy $\mathbf{E}\cdot\mathbf{B} = 0$ and $|\mathbf{E}|=|\mathbf{B}|$, from a lattice calculation using our framework.

\subsection{Form factor decomposition for the vacuum polarization tensor}
A remarkable feature of this environment is that we have two null Lorentz invariants in Minkowski space
\begin{equation}
\mathcal{F}\equiv \frac{1}{4} F_{\mu\nu}F^{\mu\nu} = 0 \,,
\end{equation}
\begin{equation}
\mathcal{G}\equiv \frac{1}{4}\tilde{F}_{\mu\nu}F^{\mu\nu} = 0 \,,
\end{equation}
which implies that the vacuum does not decay~\cite{Schwinger:1951nm}.
As a consequence, the physics of a moving particle of momentum $p$ in this background is governed by the dynamical quantum parameter
\begin{equation}
\chi^2 = \frac{e^2}{m^6}\bar{\chi}^2\,,\quad
\bar{\chi}^2\equiv(F_{\mu\nu} p^\nu)^2\,.
\end{equation}

To preserve this structure, in the Euclidean theory that we are interested in, the electric field is connected to its real, Minkowski counterpart via 
\begin{equation}
\mathbf{E}_{\rm E} = -i\mathbf{E}_{\rm M}\,.
\end{equation}
Under this choice, it is possible to have $\chi^2$ taking all values, but as usual, only the spacelike $p$ can be directly accessible from a Euclidean calculation. 

Define the Euclidean E\&M vector current
\begin{equation}
J_\mu(x)\equiv \bar{\psi}\gamma_\mu\psi\,.
\end{equation}
The vacuum polarization tensor can be calculated via
\begin{equation}\label{eq:vp}
\begin{split}
\Pi_{\mu\nu}(x;\Aext) \equiv & \frac{-1}{\mathcal{Z[\Aext]}}
\int \mathcal{D}[A,\psi,\bar{\psi}] J_\mu(x)J_\nu(0) e^{-S_{\rm g}[A] - S_{\rm f}[\Aext, A, \psi,\bar{\psi}]}\,.
\end{split}
\end{equation}

In the presence of an external potential, the background is not isotropic anymore;
therefore, the vacuum polarization tensor is characterized by three form factors $\pi_i$
\begin{equation}
\Pi_{\mu\nu}(p|\ell,\tilde{\ell}) 
\equiv \int d^4 x\; e^{ipx}\;\Pi_{\mu\nu}(x;\Aext)
= 
\sum_{i=1}^3\pi_i(p^2,\chi^2) \mathcal{P}_i(p,\ell,\tilde{\ell})\,,
\end{equation}
where
\begin{equation}
\mathcal{P}_1^{\mu\nu} \equiv g^{\mu\nu} - \frac{p^\mu p^\nu}{p^2}\,,\quad 
\mathcal{P}_2^{\mu\nu} \equiv \frac{\ell^\mu\ell^\nu}{\ell^2}\,,\quad
\mathcal{P}_3^{\mu\nu} \equiv \frac{\tilde{\ell}^\mu\tilde{\ell}^\nu}{\tilde{\ell}^2}\,,
\end{equation}
\begin{equation}
\ell^\mu\equiv F_{\mu\nu}p^\nu\,,\quad
\tilde{\ell}^\mu \equiv \tilde{F}_{\mu\nu}p^\nu\,,
\end{equation}
with
\begin{equation}
\ell^2 = \tilde{\ell}^2 = \bar{\chi}^2\,.
\end{equation}

As the UV divergence is due to the high-frequency modes of the dynamical photon, it remains the same as in the unperturbed case. 
In consequence, we can regulate the vacuum polarization tensor by first subtracting it with the one in the unperturbed theory
\begin{equation}\label{eq:vpsub}
\bar{\Pi}_{\mu\nu}(p|\ell,\tilde{\ell})\equiv \Pi_{\mu\nu}(p|\ell,\tilde{\ell}) - \Pi_{\mu\nu}^{(0)}(p)\,,\quad
\Pi_{\mu\nu}^{(0)}(p)\equiv \Pi_{\mu\nu}(p|0,0)
=p^2\Pi^{(0)}(p^2)\mathcal{P}_{1}^{\mu\nu}
\,.
\end{equation}
Here $\pi_2$ and $\pi_3$ are finite and require no further renormalization other than the vector current renormalization constant to respect the Ward identity.
We will identify them with the physical quantities $\bar{\pi}_2$ and $\bar{\pi}_3$.
As for $\pi_1$, it is renormalized after the subtraction
\begin{equation}
\bar{\pi}_1(p^2,\chi^2) \equiv \pi_1(p^2,\chi^2) - p^2\Pi^{(0)}(p^2)\,.
\end{equation}

Finally, we can obtain the renormalized form factors from the linear system
\begin{equation}\label{eq:ff-syst}
\begin{pmatrix}
\bar{\pi}_1 \\ \bar{\pi}_2 \\ \bar{\pi}_3
\end{pmatrix}
= 
\begin{pmatrix}
1   & -1 & -1  \\
-1  & 2  & 1   \\
-1  & 1  & 2     
\end{pmatrix}
\begin{pmatrix}
\mathcal{P}_1^{\mu\nu}\bar{\Pi}_{\mu\nu} \\
\mathcal{P}_2^{\mu\nu}\bar{\Pi}_{\mu\nu} \\
\mathcal{P}_3^{\mu\nu}\bar{\Pi}_{\mu\nu}
\end{pmatrix}\,.
\end{equation}

\subsection{Setup for a lattice calculation}\label{sec:lat-setup}

In a CCF setup, we can choose $\mathbf{B}_{\rm E} = -E \hat{x}_3$ and $\mathbf{E}_{\rm E} = -iE\hat{x}_2$ for the external fields in Euclidean space.
We will adopt the convention that the $\hat{x}_4$-direction is the Euclidean time and work with an external potential which depends only on the coordinate $x_2$. 

Writing Eq.~\eqref{eq:vp} in the form of Eq.~\eqref{eq:R-reg} after integrating out the fermionic degrees of freedom, we have 
\begin{equation}
\Pi_{\mu\nu}(x;\Aext) = \frac{\lla \mathcal{O}_{\mu\nu}\mathcal{V}\rra}{\lla \mathcal{V} \rra}
\end{equation}
with
\begin{equation}\label{eq:omunu}
\mathcal{O}_{\mu\nu}\equiv 
\tr\left[
\gamma_\mu S(x,0)\gamma_\nu S(0,x)
\right]
-\tr\left[
\gamma_\mu S(x,0)\right]\tr\left[\gamma_\nu S(0,x)
\right]\,,
\end{equation}
where
\begin{equation}\label{eq:prop}
S(x,y)\equiv\left(\slashed{D}+\slashed{\Aext}\right)^{-1}(x,y)\,.
\end{equation}

To use the interpolation scheme Eq.~\eqref{eq:whittaker}, it is the most convenient to perform simulation with non-compact U(1) gauge action.
The renormalization pattern is simplified if one works with lattice conserved currents exclusively;
more crucially, the subtracted Eq.~\eqref{eq:vpsub} will converge to the physical result in the continuum limit under this scheme.
Consequently, for space-like $p^2<0$, one can readily compute the three projected polarization tensors on the right-hand side of Eq.~\eqref{eq:ff-syst} by Fourier transforming the coordinate-space correlator Eq.~\eqref{eq:vp} obtained with our reweighting formalism.

However, there are still subtleties that we need to resolve before a lattice calculation is possible.
One way to incorporate $\Aext$ in the lattice propagator Eq.~\eqref{eq:prop} is by putting an extra phase factor of $\exp(a\Aext(x))$ to the link variable $U_\mu(x)= \exp(ia A(x))$.
Due to the imaginary electric field in Euclidean space, periodic boundary conditions in the $\hat{x}_2$-direction will lead to large artifact when the electron wrap around the finite box.
To cure that, one can impose Dirichlet boundary conditions in the $\hat{x}_2$-direction.
In practice, one will need to stay away from the Dirichlet boundaries while computing the Fourier transform of the correlation function to avoid distortion.
For a current-current correlator, we expect large cancellations in the exponents of the forward and backward propagating leptons which results in a fall-off of $e^{-2m_{\rm eff}\Delta x_2}$ with some effective mass $m_{\rm eff}$ and operator separation $\Delta x_2$, which results in a fast convergence in the Fourier transform. 
This behavior is studied numerically in the free-field limit in Appendix~\ref{sec:free-field}.

Another issue related to the imaginary potential is a possible crossing of zero in the spectrum of the Dirac operator in the presence of that external potential.
Such a potential zero-crossing of course depends on the dynamical photon configuration under which the Dirac operator is built. 
In the absence of dynamical photons, the Dirac operator is actually free of zero modes in the massive case, as the determinant remains unaltered from the free theory under a CCF setup~\cite{Schwinger:1951nm}.
Although exceptional configurations with near-zero modes pose severe numerical challenges, they should be suppressed by the reweighting factor, which offers a systematic way to handle them.

\section{Conclusion}\label{sec:concl}
In this work, a hybrid, reweighting-based framework allowing to deal with complex actions due to external potentials from first principles has been proposed.
In this framework, we separate the action into two parts, where the first part describes the unperturbed, positive-definite Euclidean action and the second corresponds to the difference between the target theory and the former, which introduces a complex phase to the Euclidean path-integral measure, referred to as the reweighting factor. 
Formally, the latter is a ratio of fermion determinants which in general needs to be further regulated.
In our proposal, we first generate lattice gauge configurations according to the unperturbed theory by importance sampling.
Applying gradient flow to the Monte Carlo data makes it possible to smoothly interpolate between the spacetime coordinates on the grid.
Once the lattice data embedded in the continuous spacetime, we calculate the reweighting factor from a heat-kernel-regulated worldline formalism, where the regulating proper time is taken to be the same as the gradient flow time.
The renormalization under this framework is well understood and the physical result can be obtained by extrapolating to the vanishing lattice spacing and flow time limits in turn. 

We expect the cost for the evaluation of the reweighting factor to exhibit manageable volume scaling.
Preliminary study of the scaling is given in the Proceedings of the 43rd International Symposium on Lattice Field Theory by the author~\cite{Chao:lattice2026}, based on a close-path sampling strategy similar to that proposed in Ref.~\cite{Gies:2001zp}.
Therefore, as long as the overlap between the perturbed and the unperturbed theories remains reasonable, the proposed method should provide estimates whose errors are predominantly dominated by the gauge noise at moderate numerical cost. 

A primary application of the proposed framework which has been studied in detail in this work is QED in the presence of a background electromagnetic field.
Despite the numerical complications introduced by the imaginary electric field, we have established a workflow for getting the vacuum polarization tensor in a constant crossed field background with the proposed framework.
As the Schwinger pair production rate vanishes in such a background, the vacuum free energy is not much altered when radiative corrections are included and we expect reweighting to work fine.
An immediate impact of this proposed calculation would be to verify the Ritus-Narozhnyi conjecture with a fully non-perturbative approach.

Finally, the proposed reweighting method and the renormalization procedure can be easily extended to other gauge theories with fermions.
A significant application would be to study finite chemical potential QCD at high temperatures, as no approximation of the reweighting factor is introduced with the present method, which allows to probe a larger parameter space. 
Nevertheless, as opposed to QED where a non-compact formulation gives access to the photon fields directly, the gluon fields are usually included via link variables.
As our formalism formally requires the knowledge of the gauge fields, whether it can be extracted from the corresponding link variables without having to go to larger flow time is a crucial question. 
In case of failure, as it is the Wilson loop and the field strength that are needed for the worldline formalism, direct smooth interpolations from their lattice counter parts would be a viable alternative.

\section*{Acknowledgments}
This work was supported by the U.S. Department of Energy, Office of Science, Office of Nuclear Physics under grant Contract Number DE-SC0011090.
The author would like to thank Norman Christ, William Detmold, Harvey Meyer, Arseny Mironov, Luchang Jin, Phiala Shanahan and Rui Zhang for helpful discussions. 
In particular, the author thanks Arseny Mironov for carefully reading the manuscript and pointing out mistakes in an earlier version. 
The author acknowledges the use of Google Gemini 3.1 Pro for Python code generation and mathematical calculation refinement, for which all final outputs were manually verified and remain the sole responsibility of the author.
\newpage

\appendix

\section{The Seeley-DeWitt Coefficients}\label{sec:seeley-dewitt}
We consider a local operator in $d$ dimensions of the type
\begin{equation}\label{eq:op-sdw}
\mathcal{H} = -\mathcal{D}^\mu \mathcal{D}_\mu - \mathcal{E}
\end{equation}
where $\mathcal{D}_\mu$ is a gauge-covariant derivative and $\mathcal{E}$ a potential term. 
The heat kernel associated with this operator solves the heat equation
\begin{equation}
\left(\frac{\partial}{\partial t} + \mathcal{H}^{(x)}\right) K(x, y; t) = 0\,,
\end{equation}
with the initial condition $\lim_{t \to 0} K(x, y; t) = \delta^d(x - y)\mathbb{I}$.
Note the kernel is in principle matrix-valued as it contains further degrees of freedom such as spin and color.  

To study the behavior of the kernel at small proper times, we use the following ansatz 
\begin{equation}
K(x, y; t) = \frac{1}{(4\pi t)^{d/2}} \exp\left(-\frac{|x-y|^2}{4t}\right) \sum_{k=0}^{\infty} t^k \Omega_k(x, y)\,.
\end{equation}

Substituting this expression into the heat equation, we obtain the recurrence relations
\begin{align}
(x - y)^\mu \mathcal{D}_\mu \Omega_0(x, y) &= 0\,, \\
\left(k + (x - y)^\mu \mathcal{D}_\mu\right) \Omega_k(x, y) &= - H_x \Omega_{k-1}(x, y) \quad (\text{for } k \geq 1)\,.\label{eq:recurr}
\end{align}
The Seeley-DeWitt coefficients $a_k(x)$ are defined as the coinciding limits of these functions,
\begin{equation}
a_k(x) \equiv \lim_{y \to x} \Omega_k(x, y)\,.
\end{equation}

A practical way to obtain the coefficients is to develop the gauge potential in the Fock-Schwinger gauge and take the coinciding limit.
For the study of the divergent behavior in Section~\ref{sec:sptx}, it suffices to know the first three Seeley-DeWitt coefficients, which we list here for a generic operator in the form of Eq.~\eqref{eq:op-sdw}
\begin{equation}
a_0(x) = \mathbb{I}\,,
\end{equation}
\begin{equation}
a_1(x) = \mathcal{E}(x)\,,
\end{equation}
\begin{equation}\label{eq:a2}
a_2(x) = \frac{1}{12}\mathcal{G}_{\mu\nu}\mathcal{G}^{\mu\nu} + \frac{1}{2}\mathcal{E}^2 + \frac{1}{6}\mathcal{D}^\mu \mathcal{D}_\mu \mathcal{E}\,,\quad
\mathcal{G}_{\mu\nu} \equiv [\mathcal{D}_\mu, \mathcal{D}_\nu] \,.
\end{equation}

\section{Free-field QED in a CCF background}\label{sec:free-field}

In this appendix, we study the lattice current-current correlator in free-field QED in a CCF background according to the setup described in Section~\ref{sec:lat-setup}.

We consider free Wilson fermions with an electric field $\mathbf{E}_E = -i E \hat{x}_2$ and a magnetic field $\mathbf{B}_E = -E \hat{x}_3$ in the background. 
As the system satisfies periodic boundary conditions in the $\hat{x}_1$-, $\hat{x}_3$- and $\hat{x}_4$-directions, the Dirac operator can be conveniently written in a mixed representation of $y=x_2$ and $\mathbf{p} = (p_1, p_3, p_4) \in [-\pi, \pi]^3$ after a three-dimensional Fourier transform
\begin{equation}
D_{y, y'}(\mathbf{p}) = M(\mathbf{p},y) \delta_{y, y'} - \frac{1}{2}(1 - \gamma_2)\delta_{y+1, y'} - \frac{1}{2}(1 + \gamma_2)\delta_{y-1, y'}\,,
\end{equation}
where 
\begin{equation}
\begin{split}
    M(\mathbf{p},y) \equiv & \quad (m_0 + 4)\mathbb{I} - \cos(p_1 + Ey)\mathbb{I} + i\gamma_1\sin(p_1 + Ey) \\
    & - \cos(p_3)\mathbb{I} + i\gamma_3\sin(p_3) 
     - \cosh(Ey + i p_4)\mathbb{I} + \gamma_4\sinh(Ey + i p_4)\,.
\end{split}
\end{equation}

As mentioned in Section~\ref{sec:lat-setup}, the imaginary Euclidean electric field is not compatible with periodic boundary conditions and a way to circumvent this is to implement Dirichlet boundary conditions in the $\hat{x}_2$-direction.

\begin{figure}[h!]
\includegraphics[scale=0.5]{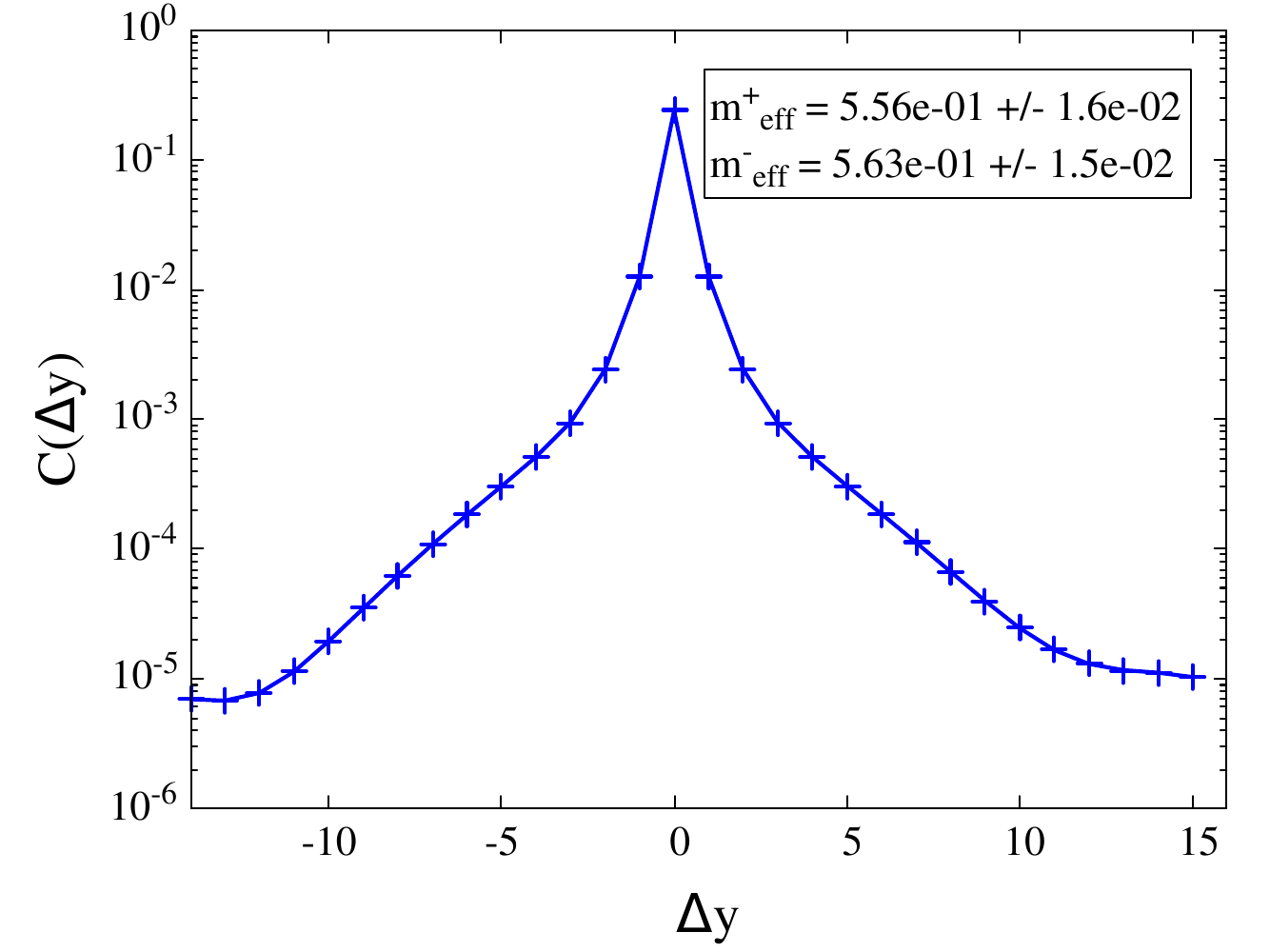}
\caption{Eq.~\eqref{eq:ccf-corr} with $E=0.2$ and $m_0= 0.02$ in lattice units. $m_{\rm eff}^+$ and $m_{\rm eff}^-$ are the effective masses fitted from the range $\Delta y \in[ 3,  8]$ and $\Delta y \in[-8, -3]$ respectively.}
\label{fig:ccf-corr}
\end{figure}

In Figure~\ref{fig:ccf-corr}, we plot the zero-momentum-projected correlator
\begin{equation}\label{eq:ccf-corr}
    C(\Delta y) = \int \frac{dp_1\; dp_3\; dp_4}{(2\pi)^3}\; \text{Tr}\left[ \gamma_\mu S(y_0 + \Delta y, y_0; \mathbf{p}) \gamma_\nu S(y_0, y_0 + \Delta y; \mathbf{p}) \right]\,,
\end{equation}
at a finite extent $N_y=32$ in the $\hat{x}_2$-direction with $y_0=0$ and $-16\leq\Delta  y \leq 15$ in lattice units.
See the caption for the parameters of the setup.
We observe effectively an asymptotic exponential fall-off at large $|\Delta y|$ which is slightly distorted by the Dirichlet boundaries.

\bibliography{references}

\end{document}